\documentclass[11pt]{article}
\begin{document}

\title{{\bf Generalized Benney Lattice and the Heavenly Equation}}
\author{A. Constandache and Ashok Das \\
Department of Physics and Astronomy,\\
University of Rochester,\\
Rochester, NY 14627-0171\\
USA\\
\\
and\\
\\
Ziemowit Popowicz \\
Institute of Theoretical Physics, \\
University of Wroc\l aw,\\
50-205 Wroclaw\\ 
Poland.}
\date{}
\maketitle

\begin{center}
{ \bf Abstract}
\end{center}

We generalize the Benney lattice and show that the new system of
equations can be reduced to a generalized Chaplygin gas as well as the
heavenly equation. We construct two infinite sets of conserved charges
and show that one of the sets can be obtained from the Lax
function. The conserved densities are related to Legendre polynomials
and we present closed form expressions for the generating functions
for these densities which also determines the Riemann invariants of
the problem. We prove that the system is bi-Hamiltonian and that the
conserved charges are in involution with respect to either of the
Hamiltonian structures. We show that the associated generalized
elastic medium equations are bi-Hamiltonian as well. We also bring out
various other interesting features of our model.

\newpage

\section{Introduction:}

The Benney lattice \cite{benney} is defined by a set of equations of the form
\begin{equation}
{\partial A_{k}\over \partial t} = A_{k+1,x} + k A_{k-1}
A_{0,x},\qquad k = 0,1,2,\cdots\label{benney} 
\end{equation}
where the subscript $x$ denotes derivative with respect to the $x$
coordinate. Several physical systems can be obtained from the Benney
lattice. In particular, note that if we identify \cite{nutku}
\begin{equation}
A_{k} = u^{k} v,\qquad k = 0,1,2,\cdots 
\end{equation}
where $u,v$ are two independent dynamical variables, then the set of
equations in (\ref{benney}) reduces to
\begin{equation}
{\partial v\over \partial t} = (uv)_{x},\quad {\partial u\over
\partial t} = uu_{x} + v_{x}
\end{equation}
which is the polytropic gas equation \cite{whitham,nutku1,das,das1,alexc}
with  $\gamma = 2$. The Benney lattice
defines a dispersionless system of equations \cite{kupershmidt} and
can be  given a Lax
description in the following manner. Consider the Lax function
\begin{equation}
L = p + \sum_{k=0}^{\infty} A_{k} p^{-(k+1)}\label{benneylax}
\end{equation}
where the coefficients, $A_{k}$, are functions of $(x,t)$ while $p$
represents the momentum of the classical phase space. Then, with the
usual canonical Poisson brackets of classical mechanics, it is easy to
verify that the Benney lattice is obtained from
\begin{equation}
{\partial L\over \partial t} = - {1\over 2}
\left\{(L^{2})_{+},L\right\}
\end{equation}
where $()_{+}$ stands for the projection with non-negative powers of
$p$. This shows that the system of equations in (1) is
integrable. Furthermore, it is also known that this system of
equations is Hamiltonian with three distinct Hamiltonian structures,
much like the system of polytropic gas equations \cite{nutku1}.

In this letter, we will generalize the Benney lattice and obtain a new
system of equations which can also be identified with various physical
systems. In particular, we will show that this system leads naturally
to a generalization of the Chaplygin gas \cite{nutku1,das1} as well as
the  heavenly
equation \cite{plebanski}. We will give a Lax description of this
system and  construct
two infinite sets of conserved charges. One of the sets of charges
follows directly from the Lax description while the connection of the
second set to the Lax function remains unclear at the
present. Interestingly, both sets of the charges are related to the
Legendre polynomials. We also
present the generating function for these charges as well as the
Riemann invariants of the system. We construct two Hamiltonian
structures  associated
with this system and show that the system is bi-Hamiltonian (is a
pencil) \cite{magri}. This is quite easily seen in a suitable choice of the
dynamical variables. We also
show the involution of the charges directly from the structure of the
generating functions. The generalized elastic medium equations,
associated with this system, are also shown to be bi-Hamiltonian. We
discuss the connection of the heavenly equation with this system of
equations and conclude with some discussions on various other aspects of
this sytem.

\section{Generalized Benney Lattice}

Let us consider a Lax function of the form
\begin{equation}
L = p^{-1} + \sum_{k=0}^{\infty} A_{k} p^{k+1}\label{lax0}
\end{equation}
If we assume the Poisson brackets of the classical phase space to be
modified as \cite{gurses}
\begin{equation}
\left\{A,B\right\}_{m} = p^{m} \left\{A,B\right\}
\end{equation}
then, it is easy to check that the Lax equation
\begin{equation}
{\partial L\over \partial t} = {1\over 2} \left\{(L^{2})_{\leq 0} ,
L\right\}_{m} 
\end{equation}
does not lead to a consistent set of equations for $m=0$ (namely, the
standard Poisson bracket relations). For $m=1$,
we have a set of consistent equations provided all the odd
coefficients, $A_{2k+1}$, in the Lax function vanish. Thus, taking our
Lax function as
\begin{equation}
L = p^{-1} + \sum_{k=0}^{\infty} A_{k} p^{2k+1}\label{lax}
\end{equation}
the Lax equation
\begin{equation}
{\partial L\over \partial t} = {1\over 2} \left\{(L^{2})_{\leq 0} ,
L\right\}_{m=1} 
\end{equation}
leads to a system of equations given by
\begin{equation}
{\partial A_{k}\over \partial t} = A_{k+1,x} + (2k+1) A_{k} A_{0,x}
\end{equation}
This system of equations can be thought of as a generalization of the
Benney lattice.

Let us note next that this system of equations can be reduced to a
generalized Chaplygin gas in the following manner. Let us identify
\begin{equation}
A_{0} = u,\quad A_{1} = - {1\over 2v^{2}},\quad A_{k} = -{1\over 2}
\sum_{i=0}^{k-1} A_{k-i-1} A_{i},\quad k > 1\label{identification}
\end{equation}
where $(u,v)$ are the independent dynamical variables. Then, it is
straightforward to show that the system of generalized
Benney lattice equations reduces to
\begin{equation}
{\partial v\over \partial t} = - (uv)_{x},\qquad {\partial u\over
\partial t} = uu_{x} + {v_{x}\over v^{3}}\label{chaplygin}
\end{equation}
This looks very much like the Chaplygin gas (polytropic gas for
$\gamma =-1$) \cite{nutku1,das1} except for the relative
sign in the two equations. It can be seen easily that there is no
transformation which will normalize the sign in both the equations.

The generalized Chaplygin equations have two infinite sets of
conserved charges which can be constructed as follows. First, let us
note that any conserved density, $h(u,v)$, of hydrodynamic type (namely, not
depending on derivatives) must satisfy
\begin{equation}
{d h\over d t} = {\partial h\over \partial u}\ u_{t} + {\partial h\over
\partial v}\ v_{t} = {d G\over dx} = {\partial G\over \partial u}\
u_{x} + {\partial G\over \partial v}\ v_{x}
\end{equation}
From the forms of the equations as well as the integrability condition
$G_{uv} = G_{vu}$, it follows that any conserved density
of hydrodynamic type associated with the
system must satisfy the relation
\begin{equation}
2u {\partial^{2} h\over \partial u\partial v} = v^{-3} {\partial^{2}
h\over \partial u\partial u} + v {\partial^{2} h\over \partial v
\partial v}\label{relation}
\end{equation}
where the conserved charges are defined as
\begin{equation}
H = \int dx\, h
\end{equation}

Let us also note from the structure of the Lax function that we can
assign the canonical dimensions $[v]=2, [u]=-2$ to the variables so
that $[uv]=0$. As a result, we can construct two independent sets of 
polynomials of the forms
\begin{equation}
h_{k} = u^{k} f_{k}(uv),\qquad \tilde{h}_{k} = v^{k} \tilde{f}_{k} (uv)
\end{equation}
They must satisfy eq. (\ref{relation}) if they are to correspond to conserved
densities of the system and it is not hard to check that this is
indeed the case provided the functions $f(uv)$ and $\tilde{f}(uv)$ are
related to the Legendre polynomials. Explicitly, $f$ and $\tilde{f}$
have to satisfy the equations
\begin{eqnarray}
z^{2} (1-z^{2}) f_{k}'' + 2z (k - (k+1)z^{2}) f_{k}' + k (k-1) f_{k} &
= & 0\nonumber\\
(1-z^{2}) \tilde{f}_{k}'' - 2z \tilde{f}_{k}' + k (k-1) \tilde{f}_{k}
& = & 0
\end{eqnarray}
where we have identified $z=uv$ and the primes denote derivatives with
respect to $z$. The second equation is easily seen to
correspond to the Legendre equation. With a change of variables, the
first equation can also be seen to be related to the Legendre equation.
The two infinite sets of
conserved densities are obtained, in this way, to be
\begin{eqnarray}
h_{k} & = & 2^{k} u^{k} {1\over (uv)^{k}}\,P_{k}(uv) = \left({v\over
2}\right)^{-k}\, P_{k} (uv)\nonumber\\
\tilde{h}_{k} & = & \left({v\over
2}\right)^{k+1}\,P_{k}(uv)\label{densities} 
\end{eqnarray}
where $k=0,1,2,\cdots $ and $P_{k}(uv)$ represent the Legendre
polynomials of order $k$. The first few densities  in the two series
have the explicit forms
\begin{eqnarray}
h_{0} & = & 1\nonumber\\
h_{1} & = & 2u\nonumber\\
h_{2} & = & 6 \left(u^{2} - {1\over 3v^{2}}\right)\label{series1}
\end{eqnarray}
and
\begin{eqnarray}
\tilde{h}_{0} & = & {v\over 2}\nonumber\\
\tilde{h}_{1} & = & {1\over 4}\,uv^{3}\nonumber\\
\tilde{h}_{2} & = & {1\over 16} \left(3u^{2}v^{5} -
v^{3}\right)\label{series2} 
\end{eqnarray}

We will show next that the first set of conserved densities in
(\ref{densities})  can be
obtained from the Lax function. Let us note that, with the
identification in eq. (\ref{identification}), we can sum the infinite
series  in (\ref{lax}) and the
Lax function can also be written in the closed form
\begin{equation}
L = p^{-1} \left(1 + 2u p^{2} + (u^{2} - {1\over v^{2}}) p^{4}\right)^{1\over
2}\label{compactlax}
\end{equation}
There are several interesting things to note from this form of the Lax
function. First of all, it is almost trivial to check from this form
that we can write
\begin{equation}
{\rm Residue}\;(p^{-1} L^{2n}) = \left({v\over 2}\right)^{-n}\,P_{n}
(uv) = h_{n}
\end{equation}
Let us note that since the Lax function in (\ref{compactlax}) involves
both  positive
and negative powers of $p$, as in the case of a non-standard Lax description of
the polytropic gas \cite{das,das1}, it is natural to expect that the
residues  calculated
around $p=0$ and $p=\infty$ may provide the two sets of conserved
densities. However, in this
case, residues calculated at both these points yield the same
conserved densities and we do not know yet how to relate the second set
of conserved densities to the Lax function. 

We also note here that the
Lax function can also be expanded around $p=\infty$ so that we can
write it in the form
\begin{equation}
L = \sum_{k=-1} A_{k} p^{-(2k+1)}
\end{equation}
which has the same form as the Lax for the Benney lattice in
(\ref{benneylax}). However, because of the difference in the classical Poisson
brackets, the two systems are inequivalent. It is also interesting to
point out here that the system of equations in (\ref{chaplygin}) has a
second  Lax
description. Consider, for example, the Lax function
\begin{equation}
L = p \left(1 + 2up^{-2} + (u^{2} - {1\over v^{2}})p^{-4}\right)^{1\over 2}
\end{equation}
Then, it can be verified in a simple manner that the Lax equation
\begin{equation}
{\partial L\over \partial t} =  {1\over 2} \left\{(L^{2})_{+} ,
L\right\}_{m=1} 
\end{equation}
also leads to the equations in (\ref{chaplygin}). However, even this second Lax
function yields only the first set of conserved densities.

Once we have the conserved densities in closed form and we know that
they are related to the Legendre polynomials, we can write generating
functions for these conserved densities in closed form. For example,
it is easy to check that
\begin{equation}
T(u,v;\lambda) = {v\over \sqrt{v^{2} - 4uv^{2}\lambda + 4\lambda^{2}}}
\end{equation}
where $\lambda$ is an arbitrary constant parameter, generates the
first set of conserved densities as
\begin{equation}
h_{n} = \left.{\partial^{n} T(u,v;\lambda)\over \partial
\lambda^{n}}\right|_{\lambda = 0}
\end{equation}
Similarly, the generating function for the second set of conserved
densities has the closed form
\begin{equation}
\tilde{T} (u,v;\tilde{\lambda}) = {v\over
\sqrt{4-4uv^{2}\tilde{\lambda} + v^{2}\tilde{\lambda}^{2}}}
\end{equation}
where $\tilde{\lambda}$ is an arbitrary constant parameter and we have
\begin{equation}
\tilde{h}_{n} = \left.{\partial^{n} \tilde{T}
(u,v;\tilde{\lambda})\over \partial
\tilde{\lambda}^{n}}\right|_{\tilde{\lambda} = 0}
\end{equation}
As in the case of the polytropic gas, we find that the quantity inside
the radical determines the Riemann invariants for the system
\cite{das,das1}.  In this
case, the two Riemann invariants have the forms
\begin{equation}
\lambda_{\pm} = {v\over 2} \left(uv \pm \sqrt{u^{2}v^{2} - 1}\right)
\end{equation}

The generalized Chaplygin gas is a bi-Hamiltonian system. We find that
the system has the following two compatible Hamiltonian structures,
\begin{eqnarray}
{\cal D}_{2} & = & {1\over 8} \left(\begin{array}{cc}
0 & \partial v (u^{2}v^{2} -1)\\
 v (u^{2}v^{2} - 1) \partial & - v (u^{2}v^{2} - 1) \partial
uv^{3} -  uv^{3} \partial v(u^{2}v^{2} - 1)
\end{array}\right)\nonumber\\
{\cal D}_{3} & = & {1\over 2} \left(\begin{array}{cc}
\partial (u^{2} - v^{-2}) + (u^{2}-v^{-2}) \partial & 2 (uv)_{x}\\
- 2 (uv)_{x} & \partial (v^{2} - u^{2}v^{4}) + (v^{2} -
u^{2}v^{4}) \partial 
\end{array}\right)\label{structures}
\end{eqnarray}
The equations (\ref{chaplygin}) can be written in the Hamiltonian form
with these structures as
\begin{equation}
\left(\begin{array}{c}
u_{t}\\
v_{t}
\end{array}\right) = {\cal D}_{3} \left(\begin{array}{c}
{\delta H_{1}\over \delta u}\\
\noalign{\vskip 4pt}%
{\delta H_{1}\over \delta v}
\end{array}\right) = {\cal D}_{2} \left(\begin{array}{c}
{\delta H_{2}\over \delta u}\\
\noalign{\vskip 4pt}%
{\delta H_{2}\over \delta v}
\end{array}\right)
\end{equation}
The compatibility of these two Hamiltonian structures can also be
checked. However, it involves a lengthy calculation in these variables
and takes a much simpler form in an alternate choice of
variables. Therefore, we will discuss this question in the next
section. However, we note
here that, unlike the polytropic gas where there are three Hamiltonian
structures, in this case, we have not been able to find the analog of
the first Hamiltonian structure.

With these two Hamiltonian structures, the involution of the conserved
charges can now be shown easily. In fact, it is much simpler to study
the involution of the generating functions for the conserved densities
themselves and show that, with respect to either of the structures,
\begin{eqnarray}
\left\{T(u,v;\lambda),T(u,v;\lambda')\right\} & = & K_{x}\nonumber\\
\left\{\tilde{T}(u,v;\tilde{\lambda}),
\tilde{T}(u,v;\tilde{\lambda}')\right\} & = & L_{x}\nonumber\\
\left\{T(u,v;\lambda),\tilde{T}(u,v;\tilde{\lambda})\right\} & = &
M_{x}
\end{eqnarray}
where $K,L,M$ are complicated (but uninteresting) polynomials. This
shows that the conserved charges (which are integrals of the
densities), are in involution and, therefore, the system is completely
integrable with respect to both the Hamiltonian structures.

We conclude this section by pointing out another interesting feature of these
equations. We know that the polytropic gas and the elastic medium
equations share the same Lax function, the same conserved charges and
both these systems are bi-Hamiltonian with the same Hamiltonian
structures \cite{das}. In the present case, we also note that if we take the
second series of conserved charges from (\ref{densities}) or
(\ref{series2}), it is easy to verify that
\begin{equation}
\left(\begin{array}{c}
u_{t}\\
v_{t}
\end{array}\right) = {1\over 8} {\cal D}_{3} \left(\begin{array}{c}
{\delta \tilde{H}_{1}\over \delta u}\\
\noalign{\vskip 4pt}%
{\delta \tilde{H}_{1}\over \delta v}
\end{array}\right) = {\cal D}_{2} \left(\begin{array}{c}
{\delta \tilde{H}_{0}\over \delta u}\\
\noalign{\vskip 4pt}%
{\delta \tilde{H}_{0}\over \delta v}
\end{array}\right) = {1\over 2} \left(\begin{array}{c}
(u^{2}v^{3} - v)_{x}\\
\noalign{\vskip 4pt}%
- (u^{3}v^{6} - u v^{4})_{x}
\end{array}\right)
\end{equation}
It is clear, therefore, that the modified elastic medium equations, in this
case, are also bi-Hamiltonian.

\section{Heavenly Equation}

Let us consider the Lax function in eq. (\ref{compactlax}) and identify
\begin{equation}
a = (u^{2} - {1\over v^{2}}),\qquad b = 2u
\end{equation}
so that we can write
\begin{equation}
L = p^{-1} (1 + b p^{2} + a p^{4})^{1\over 2}
\end{equation}
It is straightforward to check that the Lax equation,
\begin{equation}
{\partial L\over \partial t} = {1\over 2} \left\{(L^{2})_{\leq 0},
L\right\}_{m=1} 
\end{equation}
leads, in these variables, to the equations \cite{gurses}
\begin{equation}
a_{t} = a b_{x},\qquad b_{t} =  a_{x}
\end{equation}
These coupled set of first order equations can also be written as a
second order equation of the form
\begin{eqnarray}
\left(\ln a\right)_{tt} & = & a_{xx}\nonumber\\
{\rm or,}\quad X_{tt} & = & \left(e^{X}\right)_{xx}
\end{eqnarray}
where we have defined $X = \ln a$. This is known as the heavenly
equation (in $1+1$ dimensions) \cite{plebanski} and can be obtained as
a  continuum
limit of the Toda lattice. This equation appears in the study of
gravitational instantons \cite {eguchi} and can be linked to the Schr\"{o}der
equation which arises in the bootstrap models as well as in
renormalization theory \cite{hagedorn}.

The description of the dynamical equations in terms of the variables,
$a,b$, is simpler and brings out many features rather nicely. For
example, the two Hamiltonian structures (\ref{structures}) can be
written in  these
variables to be
\begin{eqnarray}
{\cal D}_{2} & = & {1\over 2} \left(\begin{array}{cc}
0 & a \partial\\
\noalign{\vskip 4pt}%
\partial a & 0
\end{array}\right)\nonumber\\
{\cal D}_{3} & = & 2 \left(\begin{array}{cc}
\partial a^{2} + a^{2} \partial & a \partial b\\
\noalign{\vskip 4pt}%
b \partial a & \partial a + a \partial
\end{array}\right)
\end{eqnarray}
We note that the first two nontrivial Hamiltonian densities of
eq. (\ref{series1})  
take the forms, in these variables, as
\begin{equation}
h_{1} = {b\over 2},\qquad h_{2} = b^{2} + 2a
\end{equation}
and it is straightforward to check that the above equations can be
written in the Hamiltonian forms
\begin{equation}
\left(\begin{array}{c}
a_{t}\\
b_{t}
\end{array}\right) = {\cal D}_{3} \left(\begin{array}{c}
{\delta H_{1}\over \delta a}\\
\noalign{\vskip 4pt}%
{\delta H_{1}\over \delta b}
\end{array}\right) = {\cal D}_{2} \left(\begin{array}{c}
{\delta H_{2}\over \delta a}\\
\noalign{\vskip 4pt}%
{\delta H_{2}\over \delta b}
\end{array}\right)
\end{equation}
It is clear that these Hamiltonian structures have a  much simpler
form  than the ones given in the last section, although they are
equivalent to these. More importantly, we note now that under a shift
$a\rightarrow a, b\rightarrow b+\lambda$, where $\lambda$ is an
arbitrary constant parameter
\begin{equation}
{\cal D}_{3} \longrightarrow {\cal D}_{3} + 4 \lambda {\cal
D}_{2}
\end{equation}
This shows that the two structures are compatible and that the system
is bi-Hamiltonian (otherwise also known as a pencil system). The
integrability of this system, therefore,
follows. This also brings out why it is more involved to explicitly
see the compatibility of the two structures in the old variables,
namely, the appropriate shift is highly nontrivial in the $(u,v)$
variables. 

Furthermore, since the Hamiltonian structures have a simpler form in
these variables, the recursion operator can also be constructed
easily. It has the form
\begin{equation}
{\cal R} = {\cal D}_{2}^{-1} {\cal D}_{3} = 4 \left(\begin{array}{cc}
b - a^{-1}\partial^{-1} ab_{x} & 2 - a^{-1}\partial^{-1} a_{x}\\
\noalign{\vskip 4pt}%
2a & b
\end{array}\right)
\end{equation}
The recursion operator, in the old variables can, of course, be
obtained from this through a coordinate redefinition. However, the
form is much more complicated and is uninteresting to list here. 

\section{Summary and Discussions}

In this letter, we have generalized the Benney lattice which can
describe a generalized Chaplygin gas equation. We have constructed the
Lax representation for this system and have constructed two infinite
sets of conserved charges which are related to the Legendre
polynomials. We can relate one of these two sets to the residues of
the Lax function and the relation of the second set to the Lax is
unclear at the present. We have given closed form expressions for the
generating functions for these densities which also leads to the two
Riemann invariants for the system. We have shown that the system is
bi-Hamiltonian (a pencil system) and have shown the involution of
charges thereby proving complete integrability of the system. We have
also shown that the associated generalized elastic medium equations
are bi-Hamiltonian as well. This system of generalized Chaplygin gas
equations can be transformed to the heavenly equation which arises in
many branches of physics. Many features of the system take a simpler
form in this description.

We would now like to make some general comments on some other aspects
of our system. First, we note that the higher order equations of the
hierarchy of generalized Chaplygin gas can be obtained from
\begin{equation}
{\partial L\over \partial t_{n}} = {1\over 2n} \left\{(L^{2n})_{\leq 0} ,
L\right\}_{m=1},\qquad n=1,2,\cdots
\end{equation}
where $L$ is the Lax function in (\ref{lax}) or
(\ref{compactlax}). However, in addition, we have also found that the
system of equations
\begin{equation}
{\partial L\over \partial t} = {1\over 2n+1} \left\{(L^{2n+1})_{\leq
0} , L\right\}_{m=2}
\end{equation}
also leads to consistent equations. However, we have not analyzed
integrability properties of such systems completely. It is also worth
noting that the Lax function in (\ref{lax0}) leads through
\begin{equation}
{\partial L\over \partial t} = {1\over 2} \left\{(L^{2})_{\leq 0} ,
L\right\}_{m=2} 
\end{equation}
to the Benney lattice. Finally, without going into details, we would
like to note that we can generalize the Lax function in
(\ref{compactlax}) to
\begin{equation}
L = p^{1-n} \left(1 + A_{1} p^{n} + A_{2} p^{2n}\right)^{1\over n}
\end{equation}
This will lead to a generalized set of equations
\begin{eqnarray}
A_{1,t} & = & {2(n-1)\over n} \left( A_{2,x} - {(n-2)\over n} A_{1}
A_{1,x}\right)\nonumber\\
A_{2,t} & = & 2\left(- {(n-3)\over n} A_{2} A_{1,x} + {(n-2)\over
n^{2}} A_{1} A_{2,x}\right)
\end{eqnarray}
following from the Lax equation
\begin{equation}
{\partial L\over \partial t} =  {1\over n} \left\{(L^{2})_{\leq 0} ,
L\right\}_{m=n-1}
\end{equation}
All these equations are likely to be integrable since they follow from
a Lax description. However, we have not studied these systems in more
detail. 

\section*{Acknowledgments}

This work was supported in part by US DOE grant
no. DE-FG-02-91ER40685 as well as by NSF-INT-0089589.

\end{document}